\documentclass[prb,aps,twocolumn,amsmath,amssymb,floatfix,
superscriptaddress]{revtex4}

\usepackage[dvips]{graphics}
\usepackage{color}
\definecolor{dred}{rgb}{0,0,0.6}

\date{\today}

\begin{document}

\title{Bias induced circular spin current: Effects of environmental 
dephasing and disorder}

\author{Moumita Patra}

\affiliation{Physics and Applied Mathematics Unit, Indian Statistical
Institute, 203 Barrackpore Trunk Road, Kolkata-700 108, India}

\author{Santanu K. Maiti}

\email{santanu.maiti@isical.ac.in}

\affiliation{Physics and Applied Mathematics Unit, Indian Statistical
Institute, 203 Barrackpore Trunk Road, Kolkata-700 108, India}

\begin{abstract}

Analogous to circular spin current in an isolated quantum loop, bias induced 
spin circular current can also be generated under certain physical 
conditions in a nanojunction having single and/or multiple loop geometries
which we propose first time, to the best of our concern, considering a 
magnetic quantum system. The key aspect of our work is the development
of a suitable theory for defining and analyzing circular spin current in 
presence of environmental dephasing and impurities. Unlike transport current 
in a conducting junction, circular current may enhance significantly in 
presence of disorder and phase randomizing processes. Our analysis provides 
a new spin dependent phenomenon, and can give important signatures in 
designing suitable spintronic devices as well as selective spin regulations.

\end{abstract}

\maketitle

\section{Introduction}

The phenomenon of bias induced circular charge current in a conducting nano 
junction having single or multiple loop geometries has been a new paradigm
of research over last few years~\cite{cir1,cir2,cir3,cir4,cir5,cir6,cir7,
cir8,cir9,cir10}. We are mostly familiar with transport current, which is 
usually referred as junction current, through a source-conductor-drain 
bridge system. But, when the bridging conductor contains a loop structure, 
a net circular current may be generated due to voltage bias satisfying some 
conditions~\cite{cir1,cir2,cir3,cir4,cir5,cir6,cir7,cir8,cir9,cir10}. This 
is quite analogous to the appearance of circular current (more usually known 
as persistent current) in an isolated mesoscopic ring-like structure (not 
connected with external baths) upon the application of magnetic 
field~\cite{pc1,pc2,pc3,pc4,pc5}. Though the phenomena are quite similar, 
the origins of these two currents are completely different. In one case it 
is due to external magnetic field and in the other case voltage bias is 
responsible. We will focus on the latter one in our present work.

The study of bond currents in different arms~\cite{cir1,cir2,cir3,cir4,cir5,
bond1,bond2} of a connected ring-like geometry essentially triggers that 
a circular current can be possible if the electrodes are attached properly 
such that the contributions from different arms do not mutually cancel with 
each other. Naturally, a possibility of tuning such current can be imagined 
by changing the junction configuration. Now what makes this phenomenon so 
special is that, this circular current induces a very large magnetic 
field~\cite{cir5,cir6,cir7,cir8,cir9} at its center as well as away (not so 
far) from the center. Because of smaller ring size, strong magnetic field, 
in some cases it may even reach to few millitesla or even tesla, will be 
induced, that can be served in many ways. The most probable application may 
be the proper regulation of electron spin or local magnetic moment, which 
can be utilized to perform different operations like storage of data, logic
functions, spin switching, spin selective electron transmission, spin based 
quantum computations, to name a few~\cite{app1,app2,app3,app4,app5,app6,app7}.

Thus, whenever we think about the tuning of a single spin or a magnetic 
moment, the application of a `local magnetic field' may be a worthy option 
for it. Few proposals have already been made~\cite{lm1,lm2} for generating 
and controlling of magnetic field locally, among them circular current induced 
magnetic field~\cite{cir5,cir6,cir7,cir8,cir9} will be the most suitable 
one, as on one hand it is very large and on the other hand its tuning is 
relatively simple rather than other propositions. So far, the phenomenon 
of `charge circular current' in nanojuctions has been 
studied~\cite{cir1,cir2,cir3,cir4,cir5,cir6,cir7,cir8,cir9,cir10}, and no one 
has explored spin dependent circular current, to the best of our knowledge, 
which might bring several salient features along this line, and thus 
probing into it is undoubtedly very essential.

In the present communication we do an in-depth analysis of circular spin 
current in a nanojuction considering a magnetic quantum ring within a 
tight-binding (TB) framework. To make the 
\begin{figure}[ht]
{\centering \resizebox*{8.75cm}{5.75cm}{\includegraphics{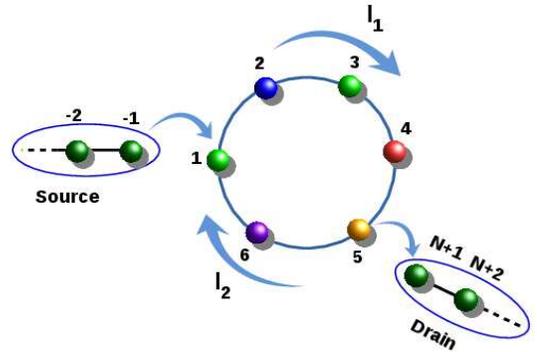}}\par}
\caption{(Color online). Schematic representation of a nanojunction having 
a single loop that may carry a net circular current in the loop, where $I_1$
and $I_2$ are the currents propagating through upper and lower arms of the 
ring, respectively.}
\label{model}
\end{figure}
model more realistic we include the effects of disorder and environmental 
dephasing. The main attention is given in developing a suitable theory for 
describing circular spin current density, and thus circular current, in 
presence of dephasing. We introduce dephasing effect by connecting B\"uttiker 
probes~\cite{dp1,dp2,dp3,dp4,dp5} at each lattice sites of the bridging 
conductor, and it can be assumed
as the most convincing and appropriate way to include phase randomization 
processes in transport phenomena. Instead of B\"uttiker probes, adding a 
constant damping factor one can also introduce dephasing into the system,
as already reported in few works~\cite{cir5,damp1,damp2}, but in this 
mechanism all the essential features may not be captured. The B\"uttiker 
probes alter the conservation conditions of different bond currents that 
should be incorporated properly to define the current densities. 

Thus, the emphasis will be given in two aspects: (i) establishing a proper
methodology for calculating circular current in presence of dephasing via
B\"uttiker probes, and (ii) defining bias induced spin circular current. 
These aspects have not been addressed earlier. We strongly believe that 
the characteristic features emerged from our analysis may provide some 
valuable inputs that can be exploited to investigate several spin dependent 
phenomena.

The arrangement of the remaining part of this paper is as follows. In 
Sec. II, we describe different spin-dependent conserved quantities and 
finite relations among them. In Sec. III, we illustrate the complete 
theoretical prescription for analyzing the phenomenon of bias induced 
circular spin current in presence of spin dependent scattering mechanism. 
In Sec. IV, we examine the accuracy of our theoretical prescription based
on which the results are computed. This will give us a confidence of our
theoretical prescription. All the essential results are throughly discussed 
in Sec. V, and finally, we summarize our findings in Sec. VI.

\section{General definition of circular current and different spin-dependent 
conserved quantities}

To define circular current, let us start with Fig.~\ref{model}, where a net
current flows from source to drain through the conducting ring. The current, 
which enters into the ring divided into two parts, (say) $I_1$ and $I_2$, and 
they re-unite at the drain end. We assign positive sign
\begin{figure}[ht]
{\centering \resizebox*{8.5cm}{3cm}{\includegraphics{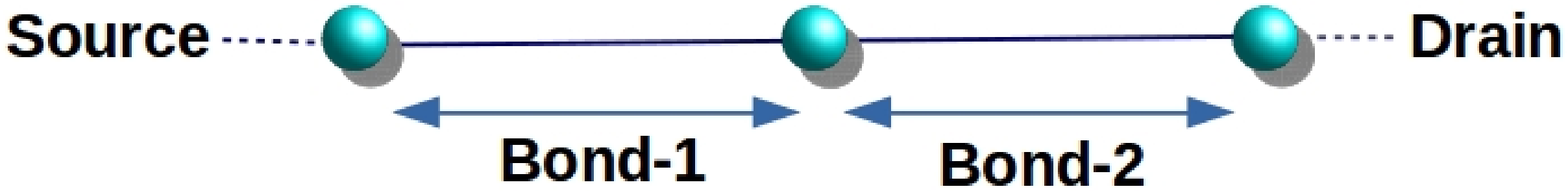}}\par}
\caption{(Color online). A nanojunction with a linear conductor having three
atomic sites.}
\label{ch}
\end{figure}
to the current propagating in the clockwise direction. If $N_1$ and $N_2$ 
be the number of atomic sites in the upper and lower arms of the ring, 
respectively, then we define circular current in the ring as~\cite{cir5,cir9}
\begin{equation}
I=\frac{I_1 N_1a + I_2 N_2a}{N_1a+N_2a}=\frac{I_1 N_1 + I_2 N_2}{N_1+N_2}
\label{1st}
\end{equation}
where, $a$ is the inter-atomic spacing. Now, for a symmetrically connected 
junction where $N_1=N_2$, the currents in the upper and lower arms are 
identical in magnitude and opposite in sign, which results a vanishing 
circular current. Thus, in order to have a net circular current, we need 
to break the status between the upper and lower arms of the 
ring~\cite{cir5,cir9}. It can be done in many ways: either by considering 
unequal lengths of a perfect junction or by introducing impurities in 
different arms of a lengthwise symmetric junction or by both.

For the calculation of currents in different sectors, first we have to
properly define the bond currents, and it is always easy to start with a 
simple linear geometry (for instance see Fig.~\ref{ch}). In this chain-like
geometry, the bond current $I_{i\rightarrow i+1}$ between any two adjacent 
sites $i$ and ($i+1$) can be expressed as~\cite{green1,green2}
\begin{equation}
I_{i\rightarrow i+1}(V) = \int J_{i \rightarrow i+1}(E) \, dE
\label{2nd}
\end{equation}
where, $J_{i \rightarrow i+1}(E)$ is the bond current density. This expression 
is equally valid for any geometry, be it a chain or any other shaped conductor. 
Now, when we stick to the liner chain model, the bond current should be exactly 
identical to that of the transport current~{\cite{cir6,damp2}}, 
defined as~\cite{green1}
\begin{equation}
I_T(V) = \frac{2e}{h} \int T(E) \, dE
\label{3rd}
\end{equation}
where, $T(E)$ is the transmission function. From Eqs.~\ref{2nd} and \ref{3rd},
we get the condition $J_{i\rightarrow i+1}=2T$ (setting $e=h=1$). The factor 
$2$ appears due to spin degeneracy. This is the fundamental 
relation to define bond current density in a linear 
geometry~\cite{cir6,damp2}, and we will extend it accordingly to calculate 
currents in different segments of any geometrical shaped conductor of our 
interest.

The scenario becomes more tricky and interesting as well when we consider
spin degree of freedom. Under this situation, the above relation becomes 
$J_{i \rightarrow i+i,\sigma}=T_{\sigma}$ where $\sigma=\uparrow, \downarrow$.
Depending on pure spin transmission and spin flip transmission, we will have
different spin dependent bond currents. Below we summarize the properties of
spin dependent bond current densities and their conservation conditions
(following the same physics as used for the spin-less case~\cite{cir6,damp2})
considering a simple set up shown in Fig.~\ref{ch} and that can be easily 
generalized for other complicated junctions as well.

\vskip 0.2cm
\noindent
$\bullet$ Case I - In absence of spin flip transmission: 

\begin{enumerate}

\item[a.] When spin flip transmission is absent, the relations between 
different spin dependent current densities with transmission components are
as follows. $J_{i \rightarrow i+1, \uparrow \uparrow}=T_{\uparrow \uparrow}$;
$J_{i \rightarrow i+1, \downarrow \downarrow}=T_{\downarrow \downarrow}$,
and $J_{i \rightarrow i+1, \uparrow \downarrow}=
J_{i \rightarrow i+1, \downarrow \uparrow}=
T_{\uparrow \downarrow}=T_{\downarrow \uparrow}=0 \, \forall \,i$.
Thus, as an example, we can write these relations for Fig.~\ref{ch} as:
$J_{1\rightarrow 2, \uparrow \uparrow}=
J_{2\rightarrow 3, \uparrow \uparrow}=T_{\uparrow \uparrow}$ and
$J_{1\rightarrow 2, \downarrow \downarrow}=
J_{2\rightarrow 3, \downarrow \downarrow}=
T_{\downarrow \downarrow}$. And, the spin flipped terms are:
$J_{1\rightarrow 2, \uparrow \downarrow}=
J_{2\rightarrow 3, \uparrow \downarrow}=
J_{1\rightarrow 2, \downarrow \uparrow}=
J_{2\rightarrow 3, \downarrow \uparrow}=
T_{\uparrow \downarrow}=T_{\downarrow \uparrow}=0$. Here, all the spin 
dependent current densities in different bonds are conserved.

\end{enumerate}

\vskip 0.2cm
\noindent
$\bullet$ Case II - In presence of spin flip transmission:

\begin{enumerate}

\item[a.] In presence of spin flip transmission, different components behave
as follows. $J_{i \rightarrow i+1, \uparrow \uparrow}\neq 
T_{\uparrow \uparrow}$;
$J_{i \rightarrow i+1, \downarrow \downarrow}\neq T_{\downarrow \downarrow}$;
$J_{i \rightarrow i+1, \uparrow \downarrow}\neq T_{\uparrow \downarrow}$;
$J_{i \rightarrow i+1, \downarrow \uparrow}\neq T_{\downarrow \uparrow} \,
\forall \,i$. Thus, for the two bonds shown in Fig.~\ref{ch} we get 
$J_{1\rightarrow 2, \uparrow \uparrow}\neq
J_{2\rightarrow 3, \uparrow \uparrow}\neq T_{\uparrow \uparrow}$;
$J_{1\rightarrow 2, \downarrow \downarrow}\neq
J_{2\rightarrow 3, \downarrow \downarrow}\neq T_{\downarrow \downarrow}$;
$J_{1\rightarrow 2, \uparrow \downarrow}\neq
J_{2\rightarrow 3, \uparrow \downarrow}\neq T_{\uparrow \downarrow}$;
and $J_{1\rightarrow 2, \downarrow \uparrow}\neq
J_{2\rightarrow 3, \downarrow \uparrow}\neq T_{\downarrow \uparrow}$. 
Here individual components are no longer conserved for different bonds.

\item[b.] Another interesting observation is that for a particular bond
$J_{i \rightarrow i+1,\uparrow\downarrow}$ becomes identical with
$J_{i \rightarrow i+1,\downarrow\uparrow}$ in that specific bond, but 
they vary from bond to bond i.e., $J_{i \rightarrow i+1,\sigma\sigma^{\prime}}$
is no longer identical with $J_{i+1 \rightarrow i+2,\sigma\sigma^{\prime}}$.

\item[c.] When we combine spin flip transmissions along with pure spin 
transmission, we get conserved quantities for each distinct bonds. They are
are prescribed as follows. Total up spin current density 
$J_{i \rightarrow i+1,\uparrow\uparrow}+
J_{i \rightarrow i+1,\downarrow\uparrow}=
T_{\uparrow\uparrow}+T_{\downarrow\uparrow} \, \forall \,i$. 
Similarly, for down spin electrons, the current density 
$J_{i \rightarrow i+1,\downarrow\downarrow}+
J_{i \rightarrow i+1,\uparrow\downarrow}=
T_{\downarrow\downarrow}+T_{\uparrow\downarrow} \, \forall \,i$.
So, for the bonds $1$ and $2$ we get the relations 
$J_{1 \rightarrow 2,\uparrow\uparrow}+
J_{1 \rightarrow 2,\downarrow\uparrow}=
J_{2 \rightarrow 3,\uparrow\uparrow}+
J_{2 \rightarrow 3,\downarrow\uparrow}=
T_{\uparrow\uparrow}+T_{\downarrow\uparrow}$
for up spin electrons, and, for down spin electrons the conditions are
$J_{1 \rightarrow 2,\downarrow\downarrow}+
J_{1 \rightarrow 2,\uparrow\downarrow}=
J_{2 \rightarrow 3,\downarrow\downarrow}+
J_{2 \rightarrow 3,\uparrow\downarrow}
=T_{\downarrow\downarrow}+T_{\uparrow\downarrow}$.

\end{enumerate}

\noindent
Here we would like to note that in the above expressions, the first term 
in the subscripts of $J$ and $T$ is used for the incident spin, while the 
second one for the transmitting electron. All these relations are equally 
valid even in the presence of disorder and environmental dephasing.

\section{Theoretical Formulation of circular spin current}

Our ultimate goal is to develop a suitable theory for 
defining spin dependent circular current in a nanojunction having a loop 
geometry in presence of impurities and environmental dephasing. To do that
we proceed in three steps. First, we try to formulate the (effective) bond
current density in presence of dephasing for the spin less case considering a
linear geometry (described in Sec. IIIA), which is always easy to understand. 
Second, we extend the idea for the same system considering spin degree of 
freedom (available in Sec. IIIB). Finally, we apply the idea into a ring-like 
geometry to have spin current density and thus spin circular current 
(discussed in Sec. IIIC).

From the conservation relations analyzed above it is clear that the
bond current densities are directly linked to the transmission functions.
Thus, to get bond currents, we need to find transmission co-efficients. 
Several methods are there like wave-guide theory~\cite{wg1,wg2,wg3,wg4}, 
transfer-matrix method~\cite{app6,tm1,tm2} and Green's function 
approach~\cite{green1,green2,new5} through which transmission probability 
can be
calculated, and in the present work, we opt the wave-guide theory based
on nearest-neighbor TB model.

\begin{figure}[ht]
{\centering \resizebox*{8cm}{3.15cm}{\includegraphics{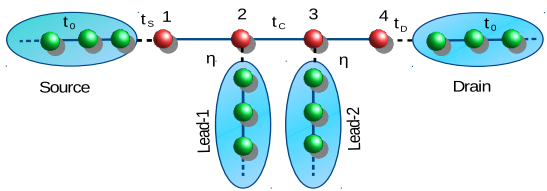}}\par}
\caption{(Color online). A nanojunction with a linear conductor in presence
of dephasing electrodes. The (dephasing) electrodes are connected at all
sites of the conductor except the end sites where source and drain electrodes
are attached.}
\label{fig1a}
\end{figure}

\vskip 0.25cm
\noindent
A. {\em Formulation of current density in a 1D chain in 
presence of dephasing for the spin less case}:
\vskip 0.25cm
Let us start with Fig.~\ref{fig1a} where a 1D non-magnetic (NM) chain
(it can also be called as channel) is coupled to source (S) and drain (D) 
electrodes along with the B\"{u}ttiker probes. All these electrodes are 
assumed to be perfect, NM and semi-infinite. The Hamiltonian for the 
entire system becomes
\begin{equation}
H= H_C + H_S + H_D + \sum_i H_B + H_T
\label{eq1}
\end{equation}
where $H_C$, $H_S$, $H_D$, and $\sum_i H_B$ represent the Hamiltonians for 
the channel (C), S, D, and the B\"uttiker probes (B), respectively. The
general form of TB Hamiltonian for these sub-systems in
nearest-neighbor hopping (NNH) approximation looks like
\begin{equation}
H_{\alpha}=\sum \epsilon_{\alpha} 
c_n^{\dagger} c_n + \sum \left(c_n^{\dagger} t_{\alpha}
c_{n-1} + c_{n-1}^{\dagger} t_{\alpha} c_n \right)
\label{eq2}
\end{equation}
where, $\alpha$=C, S, D, and B. For all the electrodes, we refer 
$t_{\alpha}=t_0$ and it is $t_C$ for the channel C. In the absence of any
voltage bias, we call $\epsilon_{\alpha}=\epsilon_0$ for the electrodes,
while for the channel C, it becomes $\epsilon_i$ ($i$ be the site index).
These site energies get modified as long as the bias is included, and the
dependence of these energies on bias can be clearly understood from our 
forthcoming analysis. The last term, $H_T$, 
of Eq.~\ref{eq1} describes the tunneling Hamiltonian due to the coupling 
of the channel with S, D, and dephasing leads, and it is also expressed 
in the usual TB form.

To calculate transmission probability and circular current density, we 
solve a set of coupled linear equations originated from the 
time-independent Schr\"odinger equation $H |\psi> = E I|\psi>$, where 
$|\psi\rangle$ represents the wave-function and $I$ is the identity matrix.
Now, for the multi-terminal set up, shown in Fig.~\ref{fig1a}, we
need to consider three different cases taking separately any one among 
the source, dephasing lead-1 and lead-2 as the incoming lead, while the 
other two including the drain as the outgoing leads. First, we consider S 
as the input lead, and thus, transmitting electrons will be collected by 
all the other leads. In this case the coupled equations for the perfect
conductor look like~\cite{damp2,wg3}:
{\small
\begin{eqnarray}
(E - \epsilon_0 - V_b)(1 + \rho_{(S)}) & = & t_{S} c_{(1,S)} +
t_0 (e^{-ik_{(S)}a} \nonumber \\
& &+ \rho_{(S)} e^{ik_{(S)}a})\nonumber \\
(E - \epsilon - V_b)c_{(1,S)} & = & t_{S} (1 + \rho_{(S)}) + t_C c_{(2,S)}
\nonumber \\
(E - \epsilon - \frac{2}{3}V_b)c_{(2,S)} & = & t_C c_{(1,S)} + t_C c_{(3,S)}
\nonumber \\
& & + \eta \tau_{(S\rightarrow B1)} e^{ik_{(B1)}a} \nonumber \\
(E - \epsilon - \frac{1}{3}V_b)c_{(3,S)} & = & t_C c_{(2,S)} + t_C c_{(4,S)}
\nonumber \\
& & + \eta \tau_{(S\rightarrow B2)} e^{ik_{(B2)}a} \nonumber \\
(E - \epsilon)c_{(4,S)} & = & t_C c_{(3,S)}\nonumber \\ 
& &+ t_{D}\tau_{(S\rightarrow D)} e^{ik_{(D)}a} \nonumber \\
(E - \epsilon_0) \tau_{(S \rightarrow D)} e^{ik_{(D)}a}
& = & t_{D} c_{(4,S)} \nonumber \\
& &+ t_0 \tau_{(S \rightarrow D)} e^{2ik_{(D)}a} \nonumber \\
(E - \epsilon_0 - \frac{2}{3}V_b) \tau_{(S \rightarrow B1)}
e^{ik_{(B1)}a} & = & \eta c_{(2,S)} \nonumber \\
& & + t_0 \tau_{(S \rightarrow B1)} e^{2ik_{(B1)}a} \nonumber \\
(E - \epsilon_0 - \frac{1}{3}V_b) \tau_{(S\rightarrow B2)} e^{ik_{(B2)}a} & = &
\eta c_{(3,S)} \nonumber \\
& & + t_0 \tau_{(S \rightarrow B2)} e^{2ik_{(B2)}a}. \nonumber \\
\label{eq3}
\end{eqnarray}}
Here we assume that, a plane wave with unit amplitude is injected from the
source end. The parameters $t_{S}$, $t_{D}$ and $\eta$ represent the 
coupling between S-to-C, C-to-D, and channel-to-dephasing lead, respectively. 
The symbols $\rho$ and $\tau$ correspond to the reflection and transmission 
amplitudes, respectively. The meaning of different equations 
and the appearance of some other factors like $V_b$, $2/3V_b$, $1/3V_b$ along 
with the wave-vectors (viz, $k_{(S)}$, $k_{(D)}$, etc.) can be understood 
explicitly as follows. To get bias induced circular current we need to apply
a finite bias across the conductor, and at the same time we have to impose the
condition that the net current through each dephasing electrode
(B\"{u}ttiker probe) is zero. To have this zero current condition for the 
dephasing electrodes, the voltages (say) $V_i$ (in $i$th electrode) should be
adjusted accordingly. These voltages ($V_i$) can easily be derived from the 
Landauer-B\"{u}ttiker current expression~\cite{green1,green2} of each virtual
lead, and noting the voltage drop at different lattice sites. 
Suppose, a bias 
$V_b$ is applied between the real electrodes i.e., $V_S=V_b$ and $V_D=0$ ($V_S$ 
and $V_D$ are the voltages associated with S and D). Then for the four-site 
chain geometry shown in Fig.~\ref{fig1a}, the voltage at site 1, where S is 
connected, will be $V_b$ and it will be zero at site 4, where D is attached.
Assuming identical voltage drop, as the voltage difference ($V_S-V_D$) is 
shared into three bonds (for 4-site chain), we can find the voltages $V_i$ at
different lattice sites of the chain. Thus at site 1, the voltage becomes $V_b$, 
and for the sites 2, 3 and 4 the voltages are $2/3V_b$, $1/3V_b$ and 0, 
respectively. Setting the bias voltage in this
fashion (viz, $2/3V_b$, $1/3V_b$) across the virtual electrodes, we can impose
the zero current condition. This prescription can easily be generalized for
any $N$-site chain as clearly demonstrated earlier by several
groups~\cite{dp3,dp4,dp5,green1,green2,dp6,niko}. In presence of non-zero bias
between S and D as the electrochemical potentials of different electrodes (real
and virtual) are different~\cite{new1}, the site energies of the electrodes 
get shifted by
constant factors associated with the voltages, and accordingly, the wave-vectors
$K_{(S)}$, $K_{(D)}$, $K_{(B1)}$ and $K_{(B2)}$ are modified. The site energies 
of the bridging conductor (viz, channel C) get modified following the bias drop 
along the chain. All these factors are incorporated properly in different site 
equations of Eq.~\ref{eq3}, and they are now voltage dependent (for 
comprehensive analysis see Refs.~\cite{new2,new3,new4}).

Solving the set of coupled equations given in Eq.~\ref{eq3}, we get the voltage
dependent transmission probability at different electronic energies at the drain 
electrode i.e., $T_{(S\rightarrow D)}$ and the bond current density between the 
sites $i$ and $i+1$ of the channel. These are respectively expressed as
\begin{equation}
T_{(S\rightarrow D)}=|\tau_{(S\rightarrow D)}|^2
\label{eq4}
\end{equation}
and
\begin{eqnarray}
J_{(i\rightarrow i+1,S)} &=& \frac{(2e/\hbar) \mbox{Im}\left[t_C\,C_{(i,S)}^*
C_{(i+1,S)} \right]}{(2e/\hbar)(1/2)t_0\sin(k_{(S)}a)} \nonumber \\
&=& \frac{2\,\mbox{Im} \left[t_C\,C_{(i,S)}^*C_{(i+1,S)} \right]}
{t_0\sin(k_{(S)}a)}.
\label{eq5}
\end{eqnarray}
The term in the denominator of Eq.~\ref{eq5} corresponds to the incident 
current density~\cite{cir5,damp2}. The subscript $S$ in the 
above current density expression is used to denote that Eq.~\ref{eq5} is
derived when the source (S) acts as the input lead. For the other input 
leads we replace $S$ by appropriate symbols as can be understood from our 
forthcoming formulation.

Now we move to the other case, where the lead-1 (B\"{u}ttiker probe) acts 
as the input lead. For this case the coupled equations are expressed as 
given in Appendix~\ref{aa}, and solving Eq.~\ref{eq6} we compute 
$T_{(B1\rightarrow D)}$ and $J_{(i\rightarrow i+1,B1)}$, like what we do 
in Eqs.~\ref{eq4} and \ref{eq5}.

Similarly, considering the other B\"{u}ttiker probe (viz, lead-2) as an input 
lead we evaluate $T_{(B2\rightarrow D)}$ and $J_{(i\rightarrow i+1,B2)}$ solving 
the set of coupled equations (Eq.~\ref{eq7}) as described in Appendix~\ref{aa}.

Following these mathematical steps we calculate the transmission 
probabilities and current densities at different segments for three different
input conditions. We ultimately want to find an effective expression of 
current density for the full system with the help of current densities of 
different regions in the presence of dephasing.

Under the condition that the current through each B\"{u}ttiker probe is zero, 
we can express the net transmission probability for the set up given in
Fig.~\ref{fig1a} as~\cite{dp3,dp4,dp5,dp6}
\begin{eqnarray}
T & = &T_{S\rightarrow D} + \sum_i T_{(i\rightarrow D)}
\frac{V_i}{V_b}\nonumber \\
& = & T_{(S\rightarrow D)} + T_{(B1\rightarrow D)}\frac{2}{3}
+ T_{(B2\rightarrow D)}\frac{1}{3}
\label{eq8}
\end{eqnarray}
where the ratio $V_i/V_b$ is determined from the above analysis. For a 
$N$-site conductor this expression will be extended accordingly.

Now, in presence of the B\"uttiker probes we define the current density
in any arbitrary bond connecting the sites $i$ and ($i+1$) as
\begin{eqnarray}
J_{(i\rightarrow i+1)} & = &\sum_{j=S,B1,B2}
J_{(i\rightarrow i+1,j)}.
\label{eq9}
\end{eqnarray}
From the current conservation conditions, we will have the following 
relations between the transmission probabilities and current densities
of different bonds of the junction configuration given in Fig.~\ref{fig1a}.
\begin{eqnarray}
2T_{(S\rightarrow D)} & = & J_{(1\rightarrow 2)}\nonumber \\
2T_{(B1\rightarrow D)} & = & J_{(2\rightarrow 3)}
-J_{(1\rightarrow 2)}\nonumber \\
2T_{(B2\rightarrow D)} & = &J_{(3\rightarrow 4)}-J_{(2\rightarrow 3)}
\label{eq10}
\end{eqnarray}
It is well known that for a strictly 1D chain $J$ should always be 
identical with $2T$~\cite{cir5,damp2} (where the factor $2$ comes due 
to spin degeneracy). Thus, combining Eqs.~\ref{eq8} and \ref{eq10}, we get
the effective expression of $J$ in presence of dephasing in the form
\begin{eqnarray}
J & = & J_{(1\rightarrow 2)} + \left(J_{(2\rightarrow 3)}
-J_{(1\rightarrow 2)}\right)\frac{2}{3} \nonumber \\
& &+ \left(J_{(3\rightarrow 4)}-J_{(2\rightarrow 3)}\right)\frac{1}{3}
\label{eq11}
\end{eqnarray}
which is utilized to find the current density in a 4-site chain as prescribed
in Fig.~\ref{fig1a}. 

The above expression of $J$ can routinely be extended for any arbitrary 1D 
chain having $N$ lattice sites where the dephasing leads are connected at all 
the sites except the boundary ones (like what is shown in Fig.~\ref{fig1a}), 
and it reads as
\begin{eqnarray}
J & = & J_{(1\rightarrow 2)}+\sum_{i=2}^{N-1} \left[J_{(i\rightarrow i+1)}
-J_{(i-1\rightarrow i)}\right]\frac{V_i}{V_b}.
\label{eq12}
\end{eqnarray}

\vskip 0.25cm
\noindent
B. {\em Formulation of current density in a 1D chain in presence of 
dephasing considering spin degree of freedom}:
\vskip 0.25cm
Now we consider the spin degree of freedom to generalize the above 
prescription for the same set up as taken in Fig.~\ref{fig1a}. Similar to 
the spin less case, here also we will have three different cases based on 
the choices of the leads among the source and two dephasing leads as 
the input one. For each input lead, we have two distinct cases depending on
which spin (up and down) of electron gets injected~\cite{wg4}. 

First we consider the source S as the input lead. We modify Eq.~\ref{eq3} 
in the spin basis to have the required sets (both for up and down spin 
incidences) of coupled equations, as prescribed in Appendix~\ref{bb}.

In the same footing, we consider the dephasing leads one by one as the 
input terminal, and modify Eqs.~\ref{eq6} and \ref{eq7}, accordingly, in
the spin basis. Solving these sets of equations we get all the required
coefficients to evaluate the spin dependent current densities at different
segments. Finally, we define the net up and down spin current densities as 
$J_{\uparrow} = J_{\uparrow\uparrow} + J_{\downarrow\uparrow}$ and
$J_{\downarrow} = J_{\uparrow\downarrow} + J_{\downarrow\downarrow}$,
respectively, where $J_{\sigma\sigma'}$ is evaluated following the 
similar kind of steps given in Eqs.~\ref{eq5}, \ref{eq9}, and \ref{eq11}.
The expressions can be generalized further for any $N$-site system following
the mechanism given in Eq.~\ref{eq12}. Using $J_{\uparrow}$ and 
$J_{\downarrow}$ we define the net charge and spin current densities as: 
$J_c=J_{\uparrow}+J_{\downarrow}$ and $J_s=J_{\uparrow}-J_{\downarrow}$, 
respectively.

\begin{figure}[ht]
{\centering \resizebox*{7cm}{6cm}{\includegraphics{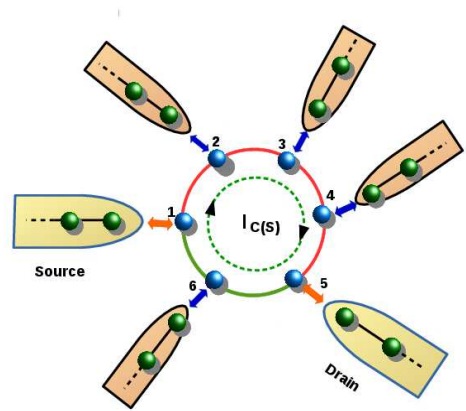}}\par}
\caption{(Color online). Ring nanojunction with dephasing electrodes, where the
red and green colors are used to indicate the upper and lower arms of the ring,
respectively. A net circular charge (spin) current ($I_{c(s)}$) is established
in presence of a finite bias $V$.}
\label{Model3}
\end{figure}

\vskip 0.25cm
\noindent
C. {\em Circular current in a 1D ring}:
\vskip 0.25cm
Utilizing the above concept, we can now determine the current density and 
circular current in a nanojunction having a ring-like geometry.
Figure~\ref{Model3} illustrates such a junction set up with dephasing 
electrodes. Like earlier, here also the dephasing leads are connected at all
the sites of the ring except the points where S and D are attached. We call 
these two points as $N_S$ and $N_D$, respectively. Let, $J_{\sigma}^u$ and 
$J_{\sigma}^l$ are the current densities in the upper and lower arms of the 
ring, respectively.
For the ring system, as two arms are there we need to consider proper weight
factors in order to calculate the circular current density. It is defined as 
\begin{equation}
J_{\sigma} =f^u J_{\sigma}^u + f^l J_{\sigma}^l
\label{eq15}
\end{equation}
where, $f^u$ and $f^l$ are the wight factors for the upper and lower arms, 
respectively. For a general $N$-site ring, these factors are:
$f^u=(N_D-1)/N$ and $f^l=(N-N_D+1)/N$ (here we fix $N_S=1$). Thus, for a
symmetrically connected ring junction $f^u=f^l=1/2$. The meaning of 
Eq.~\ref{eq15} can be simplified by looking into the set up presented in
Fig.~\ref{Model3}. Here a $6$-site ring is taken into account where S and D
are connected in such a way that the upper and lower arms contain five sites
(i.e., four bonds) and three sites (i.e., two bonds). The scheme is that,
imagine we have now two sets, where one set is a linear junction of five
sites with $N_S=1$, $N_D=5$, and the dephasing leads are attached at the sites
2, 3 and 4. On the other hand, the other set, which is also linear, contains
three sites where S and D are coupled at the two edges of the chain and the
dephasing lead is attached at the middle point. Now, both for these two sets
we determine the current densities following the steps given in the above two
sub-sections (A and B) of this section, to get $J_{\sigma}^u$ and 
$J_{\sigma}^l$. Using these current densities, we eventually calculate
$J_{\sigma}$ following the relation given in Eq.~\ref{eq15}.
Here it is important to note that the above mentioned two 
sets associated with the upper and lower arms of the ring are no longer 
decoupled with each other. When a bias is applied among S and D, an
identical voltage drop ($V_S-V_D$) takes places at the ends of the two 
arms (viz, upper and lower) of the ring as they are connected in parallel. 
For the junction configuration given in Fig.~\ref{Model3}, this voltage
(i.e., $V_S-V_D$) is shared into four bonds for the upper arm, while it
is shared into two bonds for the lower arm. Therefore, we can easily calculate
$V_i/V_b$ for the two arms following the arguments as described above 
(Sec. IIIB).

Once the spin dependent circular density is found using Eq.~\ref{eq15}, the
net circular current $I_{\sigma}$ at a bias voltage $V$ can be obtained from
the relation~\cite{cir5,cir6,cir9}
\begin{equation}
I_{\sigma}(V) = \int\limits_{E_F-\frac{eV}{2}}^{E_F+\frac{eV}{2}}
J_{\sigma}(E) \, dE
\label{eq16}
\end{equation}
where, $E_F$ is the equilibrium Fermi energy.

Finally, to check which spin dependent circular current is dominating in
a particular bias window, we can define a quantity called as circular spin
polarization as~\cite{app5} 
\begin{equation}
P = \frac{I_{\uparrow} - I_{\downarrow}}{I_{\uparrow} + I_{\downarrow}}
\label{eq17}
\end{equation}
where $P$ can be positive, or negative or even zero.

\section{Accuracy of the analysis}

Before analyzing the results, it is important to check the accuracy of our
\begin{figure}[ht]
{\centering \resizebox*{6cm}{8cm}{\includegraphics{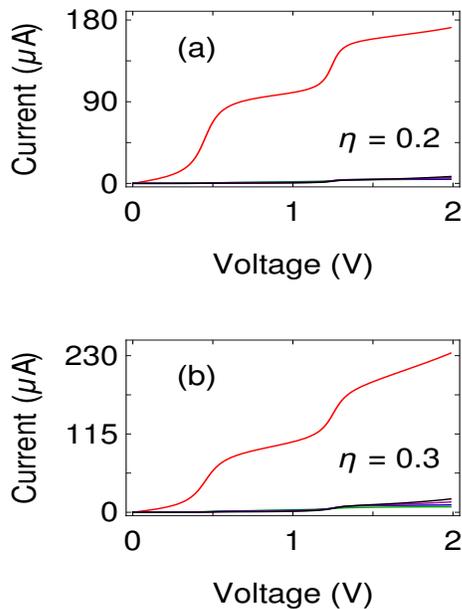}}\par}
\caption{(Color online). Currents in different B\"{u}ttiker probes together
with the drain current as a function of voltage bias for a similar kind
of conducting junction as shown in Fig.~\ref{fig1a}. Here we consider a
$6$-site chain ($N=6$), where the B\"{u}ttiker probes are connected at the
sites 2, 3, 4 and 5 sites respectively. The currents are computed for
a wide bias window setting the equilibrium Fermi energy $E_F=0$. The current
in the drain electrode is represented by the red line, while for the four
B\"{u}ttiker probes, connected at the sites 2, 3, 4 and 5 of the chain, 
the currents are described by the four different colors (green, blue, 
magenta and black) though they are not distinctly visible due to almost 
overlap of these probe currents.}
\label{accu}
\end{figure}
theoretical prescription, based on which we all the results are evaluated.
In our prescription, we assume a linear dependence of the bias drop, and 
accordingly we determine the B\"{u}ttiker voltages, though the finding 
of B\"{u}ttiker voltages is somewhat non-trivial as it is a non-linear 
problem. To validate our linear approximation, in Fig.~\ref{accu} we 
compute currents in all the B\"{u}ttiker probes together with the drain 
current, considering a $6$-site linear conductor like what is given in 
Fig.~\ref{fig1a}. Two different cases are shown depending on the strength of
$\eta$, and in both the cases clearly we see that the drain current (red 
curve) is much higher compared to the currents obtained in the B\"{u}ttiker 
probes, even at much higher voltage bias. Thus, we can emphasize that our 
approximation is quite good and can safely be utilized to analyze the bias
induced current phenomena, in presence of dephasing.

\section{Numerical Results and discussion}

We analyze the results in two parts giving the emphasis on (1) circular 
current in a ring nanojunction in presence of B\"{u}ttiker probes for 
the spin less case and then (2) extension of it in presence of spin 
dependent interaction. To explore the spin dependent phenomena, a 
clear understanding of spin independent case is definitely required.

Before starting to analyze the results, let us mention the parameter values
those are common throughout the discussion. In the absence of any voltage
bias, we set $\epsilon_0=0$, and for the perfect ring 
$\epsilon_i=0\,\forall \,i$, without loss of any generality. These site 
energies are modified in the presence of finite bias, following the 
prescription stated earlier.
The impurities in the ring are included by choosing $\epsilon_i$ in the form 
of a correlated disorder one like~\cite{aah1,aah2,aah3,aah4,aah5} 
$\epsilon_i=W\cos(2 \pi b i)$, where $W$ measures the impurity strength and 
$b$ is an irrational number. We set $b=(1+\sqrt{5})/2$ (golden mean), though 
any other irrational number can equally be taken into account. Instead of 
`correlated' disorder, one can also consider `uncorrelated' (random) site 
energies to explore the effect of disorder, but in that case we have take 
the average over a large number of distinct disordered configurations. To 
avoid it, here we ignore random distribution, and with this consideration 
no physical picture will be changed in the context of present study. $W=0$ 
corresponds to the perfect ring. The hopping integrals are: $t_0=2$, $t_C=1$ 
and $t_S=t_D=0.5$. All the energies are measured in unit of electron volt 
(eV). The system temperature and the equilibrium Fermi energy $E_F$ are 
fixed at zero. For the entire calculation we couple the source electrode 
at site $1$ of the ring (viz, $N_S=1$).

\subsubsection{In absence of spin dependent interaction}

This sub-section focuses on the characteristic properties of circular charge
current density (which sometimes may also be referred as charge current density
without always recalling the term `circular' for better readability), current
densities in different segments along with transmission probability, in a 
ring nanojunction.

\vskip 0.25cm
\noindent
A. {\em Charge current density, transmission probability and related issues}:
\vskip 0.25cm

\begin{figure}[ht]
{\centering \resizebox*{8.5cm}{7cm}{\includegraphics{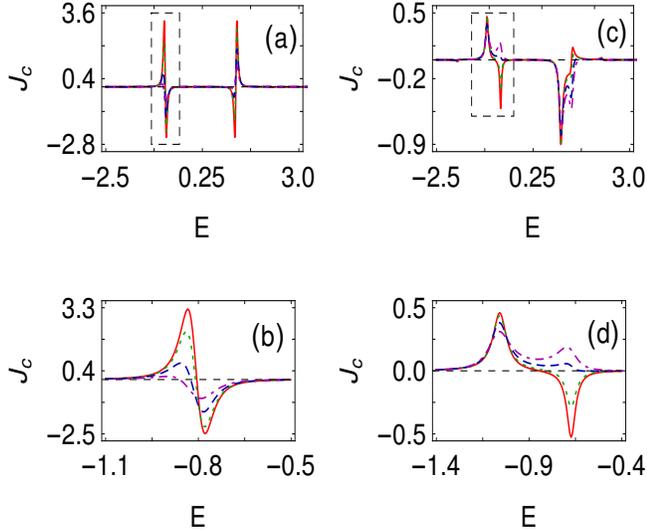}}\par}
\caption{(Color online). Charge current density $J_c$ as a function of 
energy $E$ for the (a) ordered ($W=0$) and (c) disordered ($W=0.5$) rings,
at some typical dephasing strengths ($\eta$) where the red, green, blue 
and magenta curves correspond to $\eta=0$, $0.1$, $0.2$, and $0.3$,
respectively. The enlarged version of the dashed framed regions of (a)
and (c) are shown in (b) and (d), respectively, for better viewing of the 
curves. Here we choose $N=6$, and connect the source and drain electrodes
at sites 1 and 5, respectively. Here we set the voltage bias at $0.4\,$V.}
\label{fig1}
\end{figure}

\begin{figure}[ht]
{\centering \resizebox*{8cm}{3cm}{\includegraphics{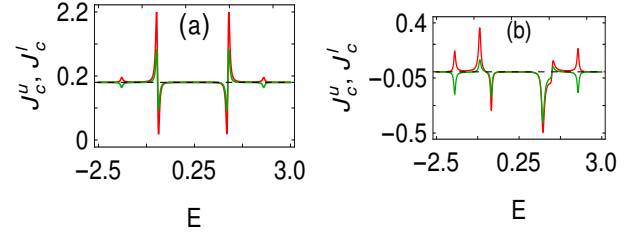}}\par}
\caption{(Color online). Charge current densities in upper (red line) and
lower (green line) arms of the ring for the (a) ordered and (b) disordered
cases, in the absence of dephasing. The other physical parameters and
ring-electrode junction configuration are kept unchanged as taken in
Fig~\ref{fig1}.}
\label{fig2}
\end{figure}

Let us start with Fig.~\ref{fig1} where the variation of $J_c$ as a function of
energy $E$ is shown both for the ordered and disordered cases at some typical
values of dephasing strength $\eta$. Several interesting features are observed, 
especially across the peaks and dips. To reveal these facts, we choose a region, 
shown by the dashed frame region, from each of the spectra given in 
Figs.~\ref{fig1}(a) and (c), and the enlarged versions of these two regions are 
placed in the bottom row of Fig.~\ref{fig1}, for better clarity of different 
colored curves. Apparently what we see from Figs.~\ref{fig1}(a) and (c) 
that four picks and dips appear in each of the spectra. All these picks and 
dips are associated with the allowed energy channels of the system.
But actually the ring junction will have six resonant energy channels since 
we set $N=6$ which yields six distinct energy eigenvalues. For the ordered 
isolated ring the eigenvalues are $-2$, $-1$, $-1$, $+1$, $+1$ and $+2$, i.e., 
the levels having eigenenergies $+1$ or $-1$ are two-fold degenerate, while 
the other two (i.e., $+2$ and $-2$) are non-degenerate. The energy channels 
associated with these eigenenergies get shifted with the inclusion of contact
leads to the ring and/or adding impurities in the ring. Thus, from the 
information of discrete energy levels of the isolated rings, the approximate
locations of the picks and dips can be estimated. A basic 
question that appears at this stage is that why no such peak or dip is observed 
at the other two energies i.e., around $\pm 2$. To explain this fact, let us 
look into the spectra given in Fig.~\ref{fig2} where the current densities in 
the upper and lower arms of the ring are shown. Here we set $\eta=0$,
and with this result the non-vanishing behavior of $J_c$ at some typical energies
for finite $\eta$ can also be understood. A tiny peak across $E=\pm 2$
appears for one arm, while a dip of almost equal strength is observed due to
the other arm at these same energies of disorder-free ring (Fig.~\ref{fig2}(a)). 
Naturally, around at $E=\pm 2$, vanishingly small contribution
in $J_c$ is obtained which is not visible in open eye from Fig.~\ref{fig1}(a).
Interestingly we see that, at the other energies the current densities for
both the two arms have identical sign ($+$ve or $-$ve), and hence it results a 
net circular current density. Identical scenario is also observed for the 
\begin{figure}[ht]
{\centering \resizebox*{8.2cm}{9.5cm}{\includegraphics{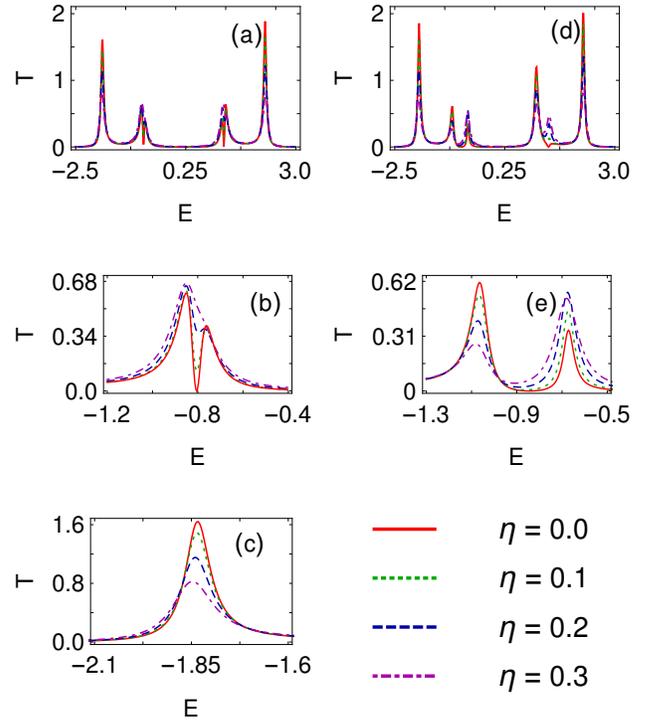}}\par}
\caption{(Color online). Transmission-energy spectra for the ordered and
disordered cases considering the identical parameter values as taken in
Fig~\ref{fig1}. The spectra shown in (b), (c) and (e) are the enlarged
version of $T$-$E$ curves within some suitable energy windows.}
\label{fig3}
\end{figure}
disordered ring, apart from an overall suppression of the charge current
densities in the arms (Fig.~\ref{fig2}(b)). This reduction is associated with
the disorder in the ring. From these results we can conclude that the sign
reversal of current densities at the two extreme energy levels remains same 
for both the ordered and disordered rings, which yields almost zero contribution 
towards $J_c$.

Now concentrate on the spectra given in Figs.~\ref{fig1}(b) and (d). It is well
known that transport current always decreases with disorder. Whereas, for the 
case of circular current the situation may be something different. The net 
circular
current for a bias voltage is obtained by integrating the current density 
$J_c(E)$, over the energy window associated with the bias. Naturally, asymmetric 
nature of $J_c$ generates more current. For the energy window shown in 
Fig.~\ref{fig1}(b) it is seen that a dip is followed by a neighboring peak and 
thus whenever we integrate over this energy window, the current will definitely 
decrease as it is the resultant contribution of picks and dips. For exactly 
equal and opposite contributions from different energy levels, the net current 
should be zero due to their mutual cancellations. An interesting feature noticed 
from Fig.~\ref{fig1}(d) is that for the disordered case there is a finite 
possibility to have phase (i.e., sign of $J_c$) reversal in presence of $\eta$, 
and thus instead of decreasing current with dephasing (as usually observed for 
the conventional transport current), one can get enhanced current since the 
successive peaks are of identical sign. Similar kind of enhancement can also be 
obtained for the perfect case, in presence of dephasing, depending on the 
junction configuration and other physical parameters, which will be understood
from our further analysis.  

The nature of peaks and dips of circular current density (i.e., magnitude
and sign) can be understood from the variation of transmission function.
The results are presented in Fig.~\ref{fig3}. Both for the ordered and
\begin{figure}[ht]
{\centering \resizebox*{7cm}{4cm}{\includegraphics{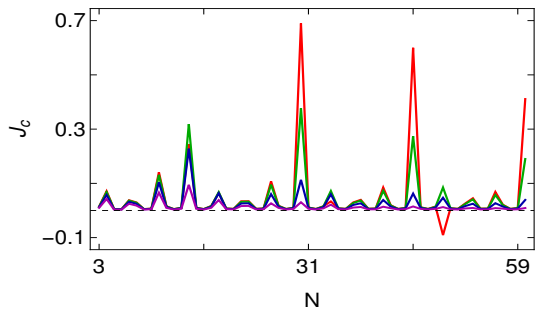}}\par}
\caption{(Color online). Dependence of $J_c$ on ring size $N$ for some
specific values of $\eta$, connecting the drain electrode at the end atomic
site ($N_D=N$) of the ring. Here we set $E=0$, $W=0$ and $V_b=0.4\,$V. The
red, green, blue and magenta lines correspond to $\eta=0$, $0.1$, $0.2,$
and $0.3$, respectively.}
\label{fig4}
\end{figure}
disordered cases large transmission appears for the two end energy levels, 
those are non-degenerate always for the isolated ring (be it ordered or 
disordered), where the peak height almost reaches to 2 (the factor 2 comes 
due to spin degeneracy). It means that at these energies transfer of electron 
is almost $100\,\%$ percent, and hence, very less contribution is obtained 
towards circular current, which is associated with the confining of electrons 
within the ring geometry.
At these resonant energies, dephasing makes a suppression of peak heights
which is clearly reflected by comparing the curves shown in Fig.~\ref{fig3}(c)
(zoomed region for a specific energy window across $E=-2$). Across the
energies $E=\sim\pm 0.8$, other many interesting features are observed and they
become more fascinating in presence of dephasing. To reveal these facts,
now focus on the spectra given in Figs.~\ref{fig3}(b) and (e), those are
zoomed versions of $T$-$E$ spectra over a selective energy window for ordered
and disordered cases, respectively. For $W=0$, antiresonance appears around at
$E=\pm 0.8$ in the absence of dephasing, where the transmission probability drops
exactly to zero. This is the generic behavior of an asymmetrically connected
interferometric geometry, and has also been been discussed in other 
contemporary works~\cite{cir5,cir6,antr}. The anti-resonant states disappear 
as long as dephasing leads are included
($\eta\neq0$), and most interesting thing is that the hight of the transmission
peaks gets increased with $\eta$ (see Fig.~\ref{fig3}(b)). This enhancement
of transmission leads to the reduction of circular current, as expected.
The situation is somewhat complicated when $W$ is finite. Looking carefully
\begin{figure}[ht]
{\centering \resizebox*{7cm}{8cm}{\includegraphics{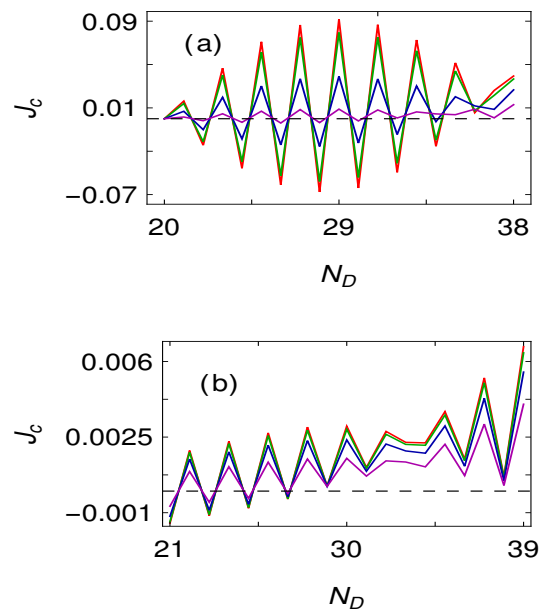}}\par}
\caption{(Color online). Effect of ring-to-drain configuration on $J_c$ at 
some typical values of $\eta$ for the bias voltage $0.4\,$V, where 
(a) $N=38$ (even) and (b) $N=39$ (odd).
Starting from the half-length of the ring, we move the drain electrode 
towards the end site of the ring. Here we fix $W=0$. The meanings of four
different colored curves are same as described in Fig.~\ref{fig4}.}
\label{fig5}
\end{figure}
into Fig.~\ref{fig3}(e), it is seen that for the two neighboring peaks the 
effect of $\eta$ is completely opposite. For one peak the hight decreases
with $\eta$, while it gets increased for the other one. This is solely
associated with the interplay between disorder and dephasing. As a result
of this there is a finite possibility to have phase reversal of $J_c$ at some
typical energies which yields higher circular current, instead of its 
conventional reduction.

\vskip 0.25cm
\noindent
B. {\em Size dependence and effect of ring-electrode interface geometry}:
\vskip 0.25cm

As quantum interference has significant impact on such properties (i.e., 
nature of circular current), it is therefore important to know how $J_c$ 
depends on the system size as well as different ring-drain configurations. 
This sub-section essentially focuses on that.

Figure~\ref{fig4} describes the dependence of $J_c$ on ring size $N$.
To have maximum contribution on $J_c$, we couple the source and drain 
electrodes in the most asymmetric configuration. A pronounced oscillation with 
$N$ is observed both for the dephasing-free ring (red curve) and the 
ring with dephasing (green, blue and magenta curves). The oscillation is 
solely associated 
with the quantum interference among electronic waves passing through different 
arms of the junction. Interestingly what we see that at lower dephasing strength
($\eta=0.1, 0.2$), $J_c$ gets much higher peak in most of the cases compared 
to the dephasing-free ring, which clearly proves that one can get much higher 
circular current in presence of dephasing and it persists up to a reasonable 
ring size.
For large enough $\eta$, as the interference effect gets reduced the overall 
envelop of $J_c$ gradually decreases which is reflected by comparing the curves 
shown in Fig.~\ref{fig4}.

Figure~\ref{fig5} describes the dependence of ring-to-drain configuration
on $J_c$. Two different cases are considered depending on the ring size $N$,
odd and even, those are presented in Figs.~\ref{fig5}(a) and (b), respectively. 
\begin{figure}[ht]
{\centering \resizebox*{8.25cm}{6.5cm}{\includegraphics{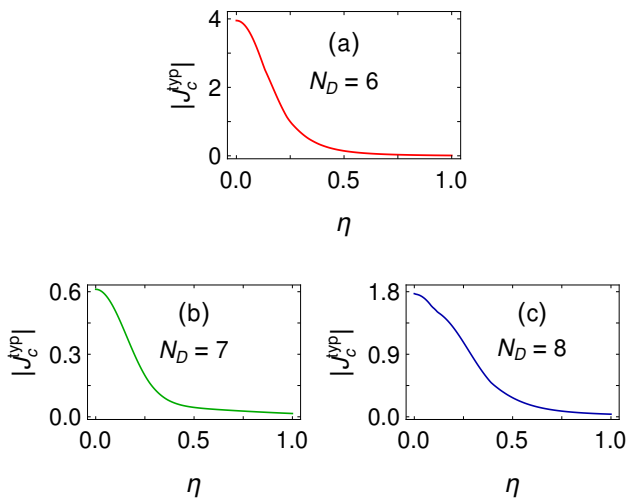}}\par}
\caption{(Color online). Variation of $|J_c^{typ}|$ as a function of dephasing 
strength for three different drain-ring positions, considering a $8$-site
perfect ring. The voltage bias is same as taken in Fig.~\ref{fig5}.}
\label{fig6}
\end{figure}
Starting from the half-length of the ring, we gradually move the drain electrode
towards the $N$th site of the ring to examine the characteristics of $J_c$. 
All these results are computed for a typical energy $E=0$, though any other 
energy can also be selected. Both for even and odd $N$, an oscillating nature 
is obtained, and the amplitude of oscillation strongly depends on the oddness 
and evenness of $N$.

So eventually what we get from Figs.~\ref{fig4} and \ref{fig5} that, $J_c$ is 
significantly influenced by quantum interference effect involving 
ring-electrode junction configuration and dephasing parameter $\eta$.

\vskip 0.25cm
\noindent
C. {\em Critical roles of $\eta$ and $W$ on circular current density}:
\vskip 0.25cm

In order to have more clear signature of $\eta$ on $J_c$ we show the dependence 
of $|J_c^{typ}|$ as a function of $\eta$ by varying it 
in a wide range, considering a ring of size $N=8$. Three different cases are 
considered depending on $N_D$, and the results are presented in Fig.~\ref{fig6}. 
$|J_c^{typ}|$ is obtained by taking the `maximum' absolute value of $J_c$ over 
the full allowed energy window. In all the three cases, the over all signature 
of $|J_c^{typ}|$-$\eta$ curve looks identical, which suggest that for large 
enough $\eta$, circular current is no longer available. This is essentially 
because of the fact that for large $\eta$ phase randomization becomes so strong 
which nullifies the effect of quantum interference. 

\begin{figure}[ht]
{\centering \resizebox*{8.5cm}{7cm}{\includegraphics{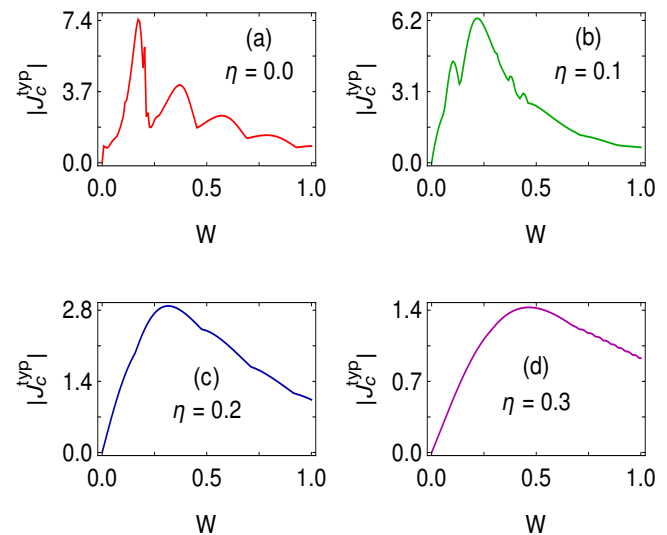}}\par}
\caption{(Color online). Role of disorder on $|J_c^{typ}|$ for a symmetrically
connected junction considering the same bias as taken in Fig.~\ref{fig5}. 
The oscillation in current density gradually decreases with $\eta$. The 
ring size is kept unchanged as taken in Fig.~\ref{fig6}.} 
\label{fig7}
\end{figure}

Finally, we concentrate on Fig.~\ref{fig7}, where the critical role of
impurities on $|J_c^{typ}|$ is shown. The interplay between dephasing ($\eta$)
and disorder ($W$) is very interesting as clearly reflected in the spectra.
For $\eta=0$, $|J_c^{typ}|$ increases suddenly and also rapidly decreases
with $W$, and it shows some irregular oscillation. With increasing $\eta$, 
the fluctuation gradually decreases and almost ceases to zero for
higher $\eta$. This is expected as quantum fluctuations get diminished with
$\eta$ because of the phase randomization. The key feature is that, here an
enhancement of current density is observed, that will provide higher circular
current, with impurity strength, which is no longer possible in the case of 
transport current (viz, the  junction current) for the conventional disordered
systems.

\subsubsection{In presence of spin dependent interaction}

Following the above analysis now we can explain the spin dependent phenomena 
and examine the critical role of dephasing, disorder, and ring-to-electrode 
junction configurations, etc., as the basic mechanisms are already discussed
for the interaction-free ring nanojunction. 

To discuss spin dependent features, we need to include spin-dependent 
scattering effect~\cite{wg4,mag1,mag2,so1,so2,so3} in the system and that 
can be done in several ways. For instance, 
by considering a magnetic quantum ring or by using a Rashba ring, or by 
some other ways. In our discussion, we concentrate on the magnetic quantum 
ring where the ring contains finite magnetic moments having strength $h_i$ 
at each lattice sites,
and their orientations can be described by the polar and azimuthal angles
$\theta_i$ and $\varphi_i$, respectively, as used in conventional polar 
co-ordinate system. Due to these magnetic sites, a spin-dependent interaction
$\vec{h}_i.\vec{\sigma}$ appears in the Hamiltonian which yields an 
effective site energy term~\cite{wg4,mag1,mag2} 
($\epsilon_i-\vec{h}_i.\vec{\sigma}$). The rest part of the Hamiltonian 
will be unchanged. Here $\vec{\sigma}$ (=$\sigma_x$, $\sigma_y$, $\sigma_z$) 
is the 
Pauli spin vector, and we assume $\sigma_z$ is diagonal. Instead of magnetic 
quantum ring, one can also use Rashba ring or a junction with other kind of 
spin-dependent scattering mechanism, and the Hamiltonian will be changed 
accordingly. Our mathematical description can be well applied for any such 
systems.

In what follows we present our results for the spin-dependent case. For this
entire section we choose $h_i=h$, $\theta_i=\theta$ and $\varphi_i=0$ for 
all sites $i$, as a matter of simplification. 

\vskip 0.25cm
\noindent
A. {\em Spin dependent circular current, circular current densities and 
spin circular current}:
\vskip 0.25cm

\begin{figure}[ht]
{\centering \resizebox*{6cm}{7.5cm}{\includegraphics{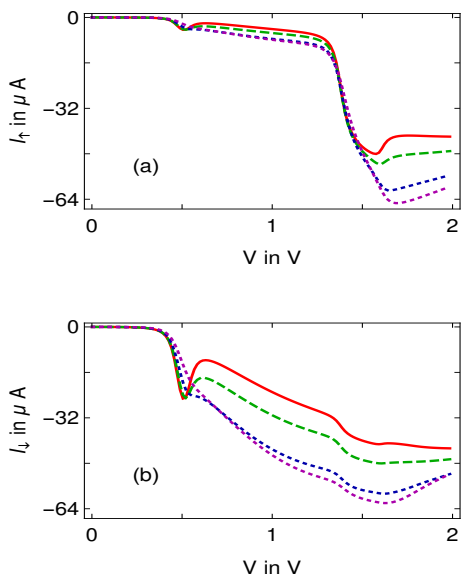}}\par}
\caption{(Color online). $I_{\sigma}$-$V$ characteristics for different
$\eta$ considering a $6$-site magnetic quantum ring with $\theta=\pi/4$, 
$h=0.5$ and $W=0$. We connect the drain at site number $5$ (i.e., $N_D=5$).
Here the red, green, blue and magenta lines correspond to $\eta=0$,
$0.1$, $0.2$ and $0.3$, respectively.}
\label{fig8}
\end{figure}

Let us start with the spin dependent circular current ($I_{\uparrow}$ and
$I_{\downarrow}$, evaluated by using Eq.~\ref{eq16}) in a perfect magnetic
quantum ring. The results are shown in Fig.~\ref{fig8} for some typical
values of dephasing strength $\eta$ considering a 6-site ring where the
drain electrode is connected at 5th site. Several noteworthy features are 
observed from the spectra Fig.~\ref{fig8}. At a first glance we see that for a 
wide bias window the current becomes zero, and the other notable thing is that
here both the increasing and decreasing natures of current with voltage can be 
obtained. This reduction of current with bias is not usually observed for 
the case of conventional transport current, which always gets enhanced, 
provided NDR effect~\cite{ndr} is not there. We explain these phenomena as 
follows. The circular current is computed by integrating the current density 
over a suitable energy window associated with finite bias voltage. For a 
specific bias when no energy level appears in the energy window no contribution 
will be there which results vanishing current. With increasing the 
bias, the energy
\begin{figure}[ht]
{\centering \resizebox*{6cm}{7.5cm}{\includegraphics{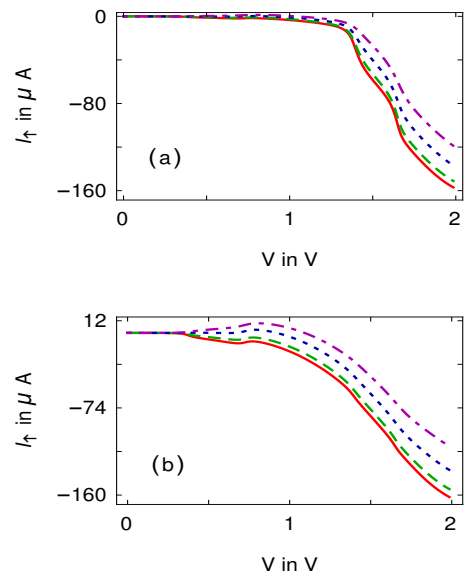}}\par}
\caption{(Color online). Same as Fig.~\ref{fig8} with $W=0.5$.}
\label{fig10}
\end{figure}
window gets wider, and now if any energy level falls within this window 
a finite current appears. When more energy levels are accommodated, all 
of them contribute and a net current is the sum of all these contributing 
energy channels, which thus can be either mutually cancelled with each other 
or may be finite one, as different energy channels are contributing current
in different directions ($+$ve and $-$ve). Here it is important to note that
unlike conventional transport current, circular current can have both positive
and negative signs depending on the contributing currents.

Along with the above facts more interesting patterns are also observed when 
we include impurities in the system. To reveal these facts look into the 
spectra given in Fig.~\ref{fig10} where the up and down spin currents are 
shown for a 6-site disordered ring setting $W=0.5$. For a wide voltage bias
currents are very small, associated with the appearance of the contributing
energy channels, and beyond that regime, current increases rapidly with 
the bias.
\begin{figure}[ht]
{\centering \resizebox*{6.5cm}{4cm}{\includegraphics{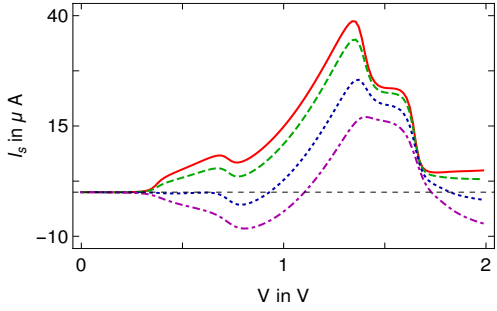}}\par}
\caption{(Color online). Dependence of spin circular current 
($I_s=I_{\uparrow}-I_{\downarrow}$) on bias voltage $V$ for the identical 
set up as taken in Fig.~\ref{fig10}. The four different colored curves have
the identical meaning as prescribed in Fig.~\ref{fig8}.}
\label{fig11}
\end{figure}
The notable thing is that for a large voltage region the magnitudes of 
both $I_{\uparrow}$ and $I_{\downarrow}$ in the disordered ring are very 
large compared to the perfect one, which can be clearly noticed by comparing
the spectra shown in Figs.~\ref{fig8} and \ref{fig10}. This is solely 
associated with the current density profile of the junction. For the ordered
case the resonant picks and dips are comparatively symmetric than the 
disordered one. More symmetric picks and dips naturally produce lesser net 
current. Thus to have higher circular current we need to have more asymmetric
current density profile. So what emerges 
from Fig.~\ref{fig10} is that, in presence of disorder higher spin dependent
current can be obtained in different voltage windows. At the same time it
is also possible to have one phase of current ($+$ve or $-$ve) for a
wide bias voltage. 

From the variations of $I_{\uparrow}$ and $I_{\downarrow}$, as given in 
Fig.~\ref{fig8} and Fig.~\ref{fig10}, an obvious question arises that how 
spin circular current $I_s(=I_{\uparrow}-I_{\downarrow})$
varies as a function of bias voltage. 
\begin{figure}[ht]
{\centering \resizebox*{6cm}{7.5cm}{\includegraphics{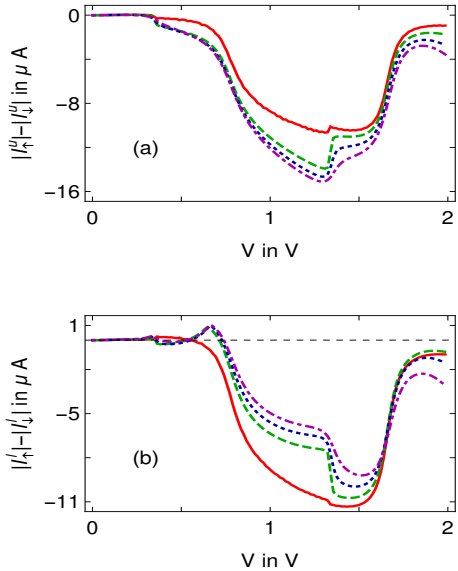}}\par}
\caption{(Color online). Difference between up and down spin currents in 
upper and lower arms of the ring as a function of bias voltage for a 
symmetrically connected $6$-site ring at different values of $\eta$. Here
we choose $W=0.5$, $h=0.5$ and $\theta=\pi/4$. Different colored curves 
represent the similar meaning as prescribed in other figures.}
\label{fig12}
\end{figure}
The answer is given in Fig.~\ref{fig11}, where we show the variation of
spin current for the same junction set up as taken in Fig.~\ref{fig10}.
The interplay between the disorder and environmental dephasing is undoubtedly
interesting. The complete phase reversal along with enhancement of spin
current can be achieved by selectively choosing the bias voltage and other
physical parameters describing the system.

From the above discussion it is now clear that a situation may happen when
in one arm the up spin current dominates the down spin current, whereas 
the phenomenon gets reversed in the other arm. The best performance can be 
achieved when the less contributing spin currents are fully suppressed, so 
that different arms will carry pure spin currents without mixing between 
$I_{\uparrow}$ and $I_{\downarrow}$. Under this
situation informations can be transferred selectively through different
segments of a nanojunction having multiple paths. Figure \ref{fig12}
describes such a possibility, where we show the variation of 
$|I_{\uparrow}| - |I_{\downarrow}|$ in the two different arms at some 
typical values of dephasing strength $\eta$, where the meaning of different
colored curves are the same as described earlier. The dephasing factor affects 
the current in different ways in the two arms. In the upper arm the magnitude 
of $|I_{\uparrow}| - |I_{\downarrow}|$ gets increased with $\eta$, while 
an opposite scenario is noticed in the other arm. These features are 
reflected in the net circular spin current.

\vskip 0.25cm
\noindent
B. {\em Polarization coefficient}:
\vskip 0.25cm

\begin{figure}[ht]
{\centering \resizebox*{8.5cm}{3.3cm}{\includegraphics{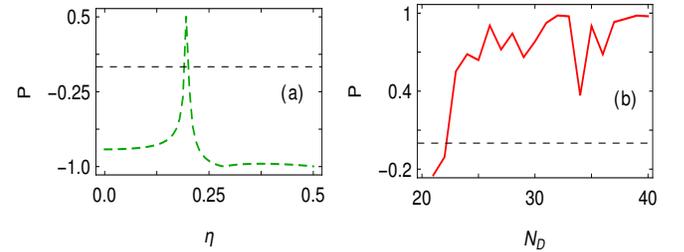}}\par}
\caption{(Color online). (a) Polarization coefficient $P$ as a function of
$\eta$. The physical parameters are: $N=6$, $W=0.5$, $h=0.5$ and 
$\theta=\pi/6$. Here we set $N_D=5$ and compute the currents at $0.5\,$V. 
(b) $P$ as a function of $N_D$ for a $40$-site ring with $\eta=0$. Here the
bias voltage is fixed at $0.25\,$V. All the other parameters are same 
as in (a).}
\label{fig13}
\end{figure}

Finally, we discuss the phenomenon of polarization coefficient $P$ that is 
calculated by using Eq.~\ref{eq17} to understand which one among $I_{\uparrow}$ 
and $I_{\downarrow}$ dominates for different input conditions. The results are 
shown in Fig.~\ref{fig13}. Let us first concentrate on Fig.~\ref{fig13}(a). It 
is clearly seen that the polarization is significantly influenced by the 
dephasing strength $\eta$. For a fixed drain position, $P$ can be changed in a 
wide range, and in some cases it may even reach to cent percent. 
At the same time, for a fixed $\eta$, a complete phase reversal of $P$ can 
also be made by altering the drain position $N_D$. Thus, the interplay between 
the environmental dephasing and ring-to-electrode junction configuration is 
extremely important to have the polarization or more precisely to characterize 
spin dependent circular currents. The role of $N_D$ is further examined by 
changing it in a wider range considering a bigger ring (see Fig.~\ref{fig13}(b)). 
The overall conclusion remains same. Under certain input conditions we can have 
net circular current completely due to one components among $I_{\uparrow}$ and 
$I_{\downarrow})$, circumventing the mixing between them.

\section{Concluding Remarks}

We have proposed a new concept of bias induced circular currents 
in a ring nanojunction in presence of impurities and environmental dephasing 
where dephasing is introduced in the form of B\"{u}ttiker probes, that may 
opens up the possibilities of designing spintronic devices and proper spin 
regulation. We have given a detailed theoretical prescription for calculating 
spin dependent current density satisfying all the conservation rules in 
presence of phase randomizing leads. Our analysis can be generalized in any 
system with any kind of spin dependent scattering. The theoretical 
prescription in presence of phase randomizing leads has not been discussed, 
even in the context of charge circular current, so far in literature to 
the best of our knowledge.

We have critically examined the characteristic features of current densities, 
branch currents, circular currents and polarization both in absence and 
presence of spin dependent interaction, and discussed thoroughly the 
interplay between the impurities and dephasing on these quantities. The 
key findings are summarized as follows.

$\bullet$ The energy levels for which electronic transmission is very high,
close to the ballistic nature, the contribution towards circular current is
too small due to less confining within the loop.

$\bullet$ Even for the disordered case, a high degree of enhancement in current
along with the phase reversal is possible with $\eta$. Though for large enough 
$\eta$ current should vanish due to complete phase randomization.

$\bullet$ The circular current is highly sensitive to the ring-electrode 
junction configuration. A pronounced oscillation has been observed.

$\bullet$ Finally, from the analysis of polarization co-efficient $P$ it can be
inferred that under certain physical conditions circular current is possible
purely due to $I_{\uparrow}$ or $I_{\downarrow}$, avoiding any mixing
between them.

\section{Acknowledgments}

SKM would like to acknowledge the financial support of DST-SERB, Government of
India, under the Project Grant No. EMR/2017/000504. SKM is also very much 
thankful to the fruitful conversation about different topics discussed in 
this work with Prof. Abraham Nitzan during the visit of SKM at Tel Aviv 
University, Israel. MP and SKM would like to thank all the reviewers for 
their constructive criticisms, valuable comments and suggestions to improve 
the quality of the work.

\appendix

\section{Coupled equations to calculate transmission probability and 
current density in absence of spin degree of freedom, considering the 
B\"{u}ttiker probes as the input leads}
\label{aa}

For the set up presented in Fig.~\ref{fig1a}, let us consider the lead-1
(B\"uttiker probe) as the input lead. Under this situation the coupled 
equations are expressed as follows.
{\small
\begin{eqnarray}
(E - \epsilon_0 - V_b) \tau_{(B1\rightarrow S)} e^{ik_{(S)}a} & = &
t_{S}c_{(1,B1)} +\nonumber \\
& &t_0 \tau_{(B1 \rightarrow S)} e^{2ik_{(S)}a}\nonumber \\
(E - \epsilon - V_b)c_{(1,B1)} & = & t_{S} \tau_{(B1\rightarrow S)}e^{ik_{(S)}a}
\nonumber \\ 
& & + t_C c_{(2,B1)} \nonumber \\
(E - \epsilon - \frac{2}{3}V_b)c_{(2,B1)} & = & t_C c_{(1,B1)} + t_C c_{(3,B1)}
\nonumber \\
& & + \eta (1 + \rho_{(B1)})\nonumber \\
(E - \epsilon - \frac{1}{3}V_b)c_{(3,B1)} & = & t_C c_{(2,B1)} + t_C c_{(4,B1)}
\nonumber \\
& & + \eta \tau_{(B1\rightarrow B2)} e^{ik_{(B2)}a} \nonumber \\
(E - \epsilon)c_{(4,B1)} & = & t_C c_{(3,B1)} + \nonumber \\
& & t_{D}\tau_{(B1\rightarrow D)} e^{ik_{(D)}a} \nonumber \\
(E - \epsilon_0) \tau_{(B1\rightarrow D)} e^{ik_{(D)}a} & = &
t_{D}c_{(4,B1)} + \nonumber \\
& &t_0 \tau_{(B1\rightarrow D)} e^{2ik_{(D)}a}\nonumber \\
(E - \epsilon_0 - \frac{2}{3}V_b)(1 + \rho_{(B1)}) & = & \eta c_{(2,B1)}+
t_0 (e^{-ik_{(B1)}a}\nonumber \\
& & + \rho_{(B1)}e^{ik_{(B1)}a}) \nonumber \\
(E - \epsilon_0 - \frac{1}{3}V_b) \tau_{(B1\rightarrow B2)} e^{ik_{(B2)}a} & = & 
\eta c_{(3,B1)} + \nonumber \\
& & t_0 \tau_{(B1\rightarrow B2)}e^{2ik_{(B2)}a}. \nonumber \\
\label{eq6}
\end{eqnarray}
}
Similarly, the coupled equations for the other B\"{u}ttiker probe 
(viz, lead-2), which can act as the input lead, are as follows.
{\small
\begin{eqnarray}
(E - \epsilon_0 - V_b) \tau_{(B2\rightarrow S)} e^{ik_{(S)}a} & = &
t_{S}c_{(1,B2)} + \nonumber \\
& & t_0 \tau_{(B2 \rightarrow S)} e^{2ik_{(S)}a}\nonumber \\
(E - \epsilon - V_b)c_{(1,B2)} & = & t_{S} \tau_{(B2\rightarrow S)} e^{ik_{(S)}a}
\nonumber \\
& &+ t_C c_{(2,B2)}\nonumber \\
(E - \epsilon- \frac{2}{3}V_b)c_{(2,B1)} & = & t_C c_{(1,B2)} + t_C c_{(3,B2)}
\nonumber \\
& & + \eta \tau_{(B2\rightarrow B1)} e^{ik_{(B1)}a} \nonumber \\
(E - \epsilon - \frac{1}{3}V_b)c_{(3,B2)} & = & t_C c_{(2,B2)} + t_C c_{(4,B4)}
\nonumber \\
& & + \eta (1 + \rho_{(B2)})\nonumber \\
(E - \epsilon)c_{(4,B2)} & = & t_C c_{(3,B2)} + \nonumber \\
& & t_{D}\tau_{(B2\rightarrow D)} e^{ik_{(D)}a} \nonumber \\
(E - \epsilon_0) \tau_{(B2\rightarrow D)} e^{ik_{(D)}a} & = & t_{D}
c_{(4,B2)} \nonumber \\
& & + t_0 \tau_{(B2\rightarrow D)} e^{2ik_{(D)}a}\nonumber \\
(E - \epsilon_0 - \frac{2}{3}V_b) \tau_{(B2\rightarrow B1)} e^{ik_{(B1)}a} & = &
\eta c_{(2,B2)} \nonumber \\
& &+ t_0 \tau_{(B2\rightarrow B1)}e^{2ik_{(B1)}a}\nonumber \\
(E - \epsilon_0 - \frac{1}{3}V_b)(1 + \rho_{(B2)}) & = & \eta c_{(3,B2)}+
t_0 (e^{-ik_{(B2)}a} \nonumber \\
& & + \rho_{(B2)}e^{ik_{(B2)}a}).
\label{eq7}
\end{eqnarray}
}
So, bond current densities between the sites $i$ and $i+1$, when the 
B\"uttiker probes $B_1$ and $B_2$ are taken as input leads, are:
\begin{eqnarray}
J_{(i\rightarrow i+1,B1)} &=& \frac{(2e/\hbar) \mbox{Im}\left[t_C\,C_{(i,B1)}^*
C_{(i+1,B1)} \right]}{(2e/\hbar)(1/2)t_0\sin(k_{(B1)}a)} \nonumber \\
&=& \frac{2\,\mbox{Im} \left[t_C\,C_{(i,B1)}^*C_{(i+1,B1)} \right]}
{t_0\sin(k_{(B1)}a)}
\label{eq7a}
\end{eqnarray}
and, 
\begin{eqnarray}
J_{(i\rightarrow i+1,B2)} &=& \frac{(2e/\hbar) \mbox{Im}\left[t_C\,C_{(i,B2)}^*
C_{(i+1,B2)} \right]}{(2e/\hbar)(1/2)t_0\sin(k_{(B2)}a)} \nonumber \\
&=& \frac{2\,\mbox{Im} \left[t_C\,C_{(i,B2)}^*C_{(i+1,B2)} \right]}
{t_0\sin(k_{(B2)}a)}
\label{eq7b}
\end{eqnarray}
respectively.

\section{Coupled equations to calculate transmission probability and 
current density in presence of spin degree of freedom, considering the 
source as the input lead}
\label{bb}

For the junction configuration given in Fig.~\ref{fig1a}, the set of coupled
equations, in presence of spin degree of freedom, considering source S as 
the input lead where electrons with up spin are injected, can be expressed
as follows.
\begin{widetext}
{\small
\begin{eqnarray}
\left[\left(\begin{array}{cc}
	E & 0 \\
	0 & E
\end{array}\right) - \left(\begin{array}{cc}
    \epsilon_0 + V_b & 0 \\
    0 & \epsilon_0 + V_b
\end{array}\right)\right]\left(\begin{array}{cc}
        1 + \rho_{\uparrow\uparrow(S)} \\
        \rho_{\uparrow\downarrow(S)}
    \end{array}\right) = \left(\begin{array}{cc}
    t_0 & 0 \\
    0 & t_0
\end{array}\right) \left(\begin{array}{cc}
    e^{-ika} + \rho_{\uparrow\uparrow(S)}e^{ik_{(S)}a} \\
	\rho_{\uparrow\downarrow(S)}e^{ik_{(S)}a}
    \end{array}\right) + \left(\begin{array}{cc}
   t_{S}  & 0 \\
    0 & t_{S}
\end{array}\right) \left(\begin{array}{cc}
   c_{\uparrow\uparrow(1,S)}  & 0 \\
    0 & c_{\uparrow\downarrow(1,S)}
\end{array}\right)\nonumber \\
\left[\left(\begin{array}{cc}
    E & 0 \\
    0 & E
\end{array}\right) - \left(\begin{array}{cc}
    \epsilon + V_b & 0 \\
    0 & \epsilon + V_b
\end{array}\right)\right] \left(\begin{array}{cc}
   c_{\uparrow\uparrow(1,S)}  & 0 \\ 
    0 & c_{\uparrow\downarrow(1,S)}
\end{array}\right) = \left(\begin{array}{cc}
    t_{S} & 0 \\ 
    0 & t_{S} 
\end{array}\right)\left(\begin{array}{cc}
    1 + \rho_{\uparrow\uparrow(S)} \\
    \rho_{\uparrow\downarrow(S)} 
    \end{array}\right) + \left(\begin{array}{cc}
    t_{C} & 0 \\ 
    0 & t_{C} 
\end{array}\right) \left(\begin{array}{cc}
   c_{\uparrow\uparrow(2,S)}  & 0 \\ 
    0 & c_{\uparrow\downarrow(2,S)}
\end{array}\right)
\nonumber \\
\left[\left(\begin{array}{cc}
    E & 0 \\ 
    0 & E 
\end{array}\right) - \left(\begin{array}{cc}
	\epsilon + \frac{2}{3}V_b & 0 \\  
    0 & \epsilon + \frac{2}{3}V_b
\end{array}\right)\right] \left(\begin{array}{cc}
   c_{\uparrow\uparrow(2,S)}  & 0 \\ 
    0 & c_{\uparrow\downarrow(2,S)}
\end{array}\right) = \left(\begin{array}{cc}
    t_{C} & 0 \\ 
    0 & t_{C}
\end{array}\right) \left(\begin{array}{cc}
   c_{\uparrow\uparrow(1,S)}  & 0 \\ 
    0 & c_{\uparrow\downarrow(1,S)}
\end{array}\right) + \left(\begin{array}{cc}
    t_{C} & 0 \\ 
    0 & t_{C} 
\end{array}\right) \left(\begin{array}{cc}
   c_{\uparrow\uparrow(3,S)}  & 0 \\ 
    0 & c_{\uparrow\downarrow(3,S)}
\end{array}\right)\nonumber \\
 + \left(\begin{array}{cc}
   \eta & 0 \\ 
    0 & \eta
\end{array}\right) \left(\begin{array}{cc}
	\tau_{\uparrow\uparrow(S\rightarrow B1)}e^{ik_{(B1)}a} \\
    \tau_{\uparrow\downarrow(S\rightarrow B1)}e^{ik_{(B1)}a} 
    \end{array}\right) \nonumber \\
\left[\left(\begin{array}{cc}
    E & 0 \\ 
    0 & E 
\end{array}\right) - \left(\begin{array}{cc}
    \epsilon + \frac{1}{3}V_b & 0 \\  
    0 & \epsilon + \frac{1}{3}V_b
\end{array}\right)\right] \left(\begin{array}{cc}
   c_{\uparrow\uparrow(3,S)}  & 0 \\ 
    0 & c_{\uparrow\downarrow(3,S)}
\end{array}\right) = \left(\begin{array}{cc}
    t_{C} & 0 \\ 
    0 & t_{C} 
\end{array}\right) \left(\begin{array}{cc}
   c_{\uparrow\uparrow(2,S)}  & 0 \\ 
    0 & c_{\uparrow\downarrow(2,S)}
\end{array}\right) + \left(\begin{array}{cc}
    t_{C} & 0 \\ 
    0 & t_{C} 
\end{array}\right) \left(\begin{array}{cc}
   c_{\uparrow\uparrow(4,S)}  & 0 \\ 
    0 & c_{\uparrow\downarrow(4,S)}
\end{array}\right) \nonumber \\
 + \left(\begin{array}{cc}
   \eta & 0 \\ 
    0 & \eta
\end{array}\right) \left(\begin{array}{cc}
	\tau_{\uparrow\uparrow(S\rightarrow B2)}e^{ik_{(B2)}a} \\
	\tau_{\uparrow\downarrow(S\rightarrow B2)}e^{ik_{(B2)}a} 
    \end{array}\right)\nonumber \\
\left[\left(\begin{array}{cc}
    E & 0 \\ 
    0 & E 
\end{array}\right) - \left(\begin{array}{cc}
    \epsilon & 0 \\  
    0 & \epsilon 
\end{array}\right)\right] \left(\begin{array}{cc}
   c_{\uparrow\uparrow(4,S)}  & 0 \\ 
    0 & c_{\uparrow\downarrow(4,S)}
\end{array}\right)
  = \left(\begin{array}{cc}
    t_{C} & 0 \\ 
    0 & t_{C}
\end{array}\right) \left(\begin{array}{cc}
   c_{\uparrow\uparrow(3,S)}  & 0 \\ 
    0 & c_{\uparrow\downarrow(3,S)}
\end{array}\right) +  \left(\begin{array}{cc}
    t_{D} & 0 \\ 
    0 & t_{D} 
\end{array}\right) \left(\begin{array}{cc}
    \tau_{\uparrow\uparrow(S\rightarrow D)}e^{ik_{(D)}a} \\
    \tau_{\uparrow\downarrow(S\rightarrow D)}e^{ik_{(D)}a} 
    \end{array}\right)\nonumber \\
\left[\left(\begin{array}{cc}
    E & 0 \\ 
    0 & E 
\end{array}\right) - \left(\begin{array}{cc}
    \epsilon_0 & 0 \\ 
    0 & \epsilon_0 
\end{array}\right)\right]
\left(\begin{array}{cc}
    \tau_{\uparrow\uparrow(S\rightarrow D)}e^{ik_{(D)}a} \\
    \tau_{\uparrow\downarrow(S\rightarrow D)}e^{ik_{(D)}a} 
    \end{array}\right) = \left(\begin{array}{cc} 
    t_{D} & 0 \\ 
    0 & t_{D} 
\end{array}\right) \left(\begin{array}{cc}
   c_{\uparrow\uparrow(4,S)}  & 0 \\ 
    0 & c_{\uparrow\downarrow(4,S)}
\end{array}\right) +
 \left(\begin{array}{cc}
    t_{0} & 0 \\
0 & t_{0}
\end{array}\right) \left(\begin{array}{cc}
    \tau_{\uparrow\uparrow(S\rightarrow D)}e^{2ik_{(D)}a} \\
    \tau_{\uparrow\downarrow(S\rightarrow D)}e^{2ik_{(D)}a}
    \end{array}\right)\nonumber \\
\left[\left(\begin{array}{cc}
    E & 0 \\
    0 & E
\end{array}\right) - \left(\begin{array}{cc}
    \epsilon_0 + \frac{2}{3}V_b & 0 \\
    0 & \epsilon_0 + \frac{2}{3}V_b
\end{array}\right)\right] \left(\begin{array}{cc}
	\tau_{\uparrow\uparrow(S\rightarrow B1)}e^{ik_{(B1)}a} \\
    \tau_{\uparrow\downarrow(S\rightarrow B1)}e^{ik_{(B1)}a}
    \end{array}\right)
 = \left(\begin{array}{cc}
   \eta & 0 \\
    0 & \eta
\end{array}\right) \left(\begin{array}{cc}
   c_{\uparrow\uparrow(2,S)}  & 0 \\
    0 & c_{\uparrow\downarrow(2,S)}
\end{array}\right)+ \left(\begin{array}{cc}
    t_0 & 0 \\
    0 & t_0
\end{array}\right) \left(\begin{array}{cc}
    \tau_{\uparrow\uparrow(S\rightarrow B1)}e^{2ik_{(B1)}a} \\
    \tau_{\uparrow\downarrow(S\rightarrow B1)}e^{2ik_{(B1)}a}
    \end{array}\right)\nonumber \\
\left[\left(\begin{array}{cc}
    E & 0 \\
    0 & E
\end{array}\right) - \left(\begin{array}{cc}
    \epsilon_0 + \frac{1}{3}V_b & 0 \\
    0 & \epsilon_0 + \frac{1}{3}V_b
\end{array}\right)\right] \left(\begin{array}{cc}
    \tau_{\uparrow\uparrow(S\rightarrow B2)}e^{ik_{(B2)}a} \\
    \tau_{\uparrow\downarrow(S\rightarrow B2)}e^{ik_{(B2)}a}
    \end{array}\right)
 = \left(\begin{array}{cc}
   \eta & 0 \\
    0 & \eta
\end{array}\right) \left(\begin{array}{cc}
   c_{\uparrow\uparrow(3,S)}  & 0 \\
    0 & c_{\uparrow\downarrow(3,S)}
\end{array}\right) + \left(\begin{array}{cc}
    t_0 & 0 \\
    0 & t_0
\end{array}\right) \left(\begin{array}{cc}
    \tau_{\uparrow\uparrow(S\rightarrow B2)}e^{2ik_{(B2)}a} \\
    \tau_{\uparrow\downarrow(S\rightarrow B2)}e^{2ik_{(B2)}a}
    \end{array}\right). \nonumber \\
\label{eq13}
\end{eqnarray}}
\end{widetext}
Similarly, for the down spin incidence the equations are:
\begin{widetext}
{\small
\begin{eqnarray}
\left[\left(\begin{array}{cc}
    E & 0 \\
    0 & E
\end{array}\right) - \left(\begin{array}{cc}
    \epsilon_0 + V_b & 0 \\
    0 & \epsilon_0 + V_b
\end{array}\right)\right]\left(\begin{array}{cc}
        \rho_{\downarrow\uparrow(S)} \\
        1 + \rho_{\downarrow\downarrow(S)}
    \end{array}\right) = \left(\begin{array}{cc}
    t_0 & 0 \\
    0 & t_0
\end{array}\right) \left(\begin{array}{cc}
    \rho_{\downarrow\uparrow(S)}e^{ik_{(S)}a} \\
    e^{-ika} + \rho_{\downarrow\downarrow(S)}e^{ik_{(S)}a} \\
    \end{array}\right) + \left(\begin{array}{cc}
   t_{S}  & 0 \\
    0 & t_{S}
\end{array}\right) \left(\begin{array}{cc}
   c_{\downarrow\uparrow(1,S)}  & 0 \\
    0 & c_{\downarrow\downarrow(1,S)}
\end{array}\right)\nonumber \\
\left[\left(\begin{array}{cc}
    E & 0 \\
    0 & E
\end{array}\right) - \left(\begin{array}{cc}
    \epsilon + V_b & 0 \\
    0 & \epsilon + V_b
\end{array}\right)\right] \left(\begin{array}{cc}
   c_{\downarrow\uparrow(1,S)}  & 0 \\
0 & c_{\downarrow\downarrow(1,S)}
\end{array}\right) = \left(\begin{array}{cc}
    t_{S} & 0 \\ 
    0 & t_{S} 
\end{array}\right)\left(\begin{array}{cc}
    \rho_{\downarrow\uparrow(S)}\\
    1 + \rho_{\downarrow\downarrow(S)}
    \end{array}\right) + \left(\begin{array}{cc}
    t_{C} & 0 \\ 
    0 & t_{C} 
\end{array}\right) \left(\begin{array}{cc}
   c_{\downarrow\uparrow(2,S)}  & 0 \\ 
    0 & c_{\downarrow\downarrow(2,S)}
\end{array}\right)
\nonumber \\
\left[\left(\begin{array}{cc}
    E & 0 \\ 
    0 & E 
\end{array}\right) - \left(\begin{array}{cc}
	\epsilon + \frac{2}{3}V_b & 0 \\  
    0 & \epsilon + \frac{2}{3}V_b
\end{array}\right)\right] \left(\begin{array}{cc}
   c_{\downarrow\uparrow(2,S)}  & 0 \\ 
    0 & c_{\downarrow\downarrow(2,S)}
\end{array}\right) = \left(\begin{array}{cc}
    t_{C} & 0 \\ 
    0 & t_{C} 
\end{array}\right) \left(\begin{array}{cc}
   c_{\downarrow\uparrow(1,S)}  & 0 \\
0 & c_{\downarrow\downarrow(1,S)}
\end{array}\right) + \left(\begin{array}{cc}
    t_{C} & 0 \\ 
    0 & t_{C} 
\end{array}\right) \left(\begin{array}{cc}
   c_{\downarrow\uparrow(3,S)}  & 0 \\ 
    0 & c_{\downarrow\downarrow(3,S)}
\end{array}\right)\nonumber \\
 + \left(\begin{array}{cc}
   \eta & 0 \\ 
    0 & \eta
\end{array}\right) \left(\begin{array}{cc}
    \tau_{\downarrow\uparrow(S\rightarrow B1)}e^{ik_{(B1)}a} \\
    \tau_{\downarrow\downarrow(S\rightarrow B1)}e^{ik_{(B1)}a} 
    \end{array}\right) \nonumber \\
\left[\left(\begin{array}{cc}
    E & 0 \\ 
    0 & E 
\end{array}\right) - \left(\begin{array}{cc}
    \epsilon + \frac{1}{3}V_b & 0 \\  
    0 & \epsilon  + \frac{1}{3}V_b
\end{array}\right)\right] \left(\begin{array}{cc}
   c_{\downarrow\uparrow(3,S)}  & 0 \\ 
    0 & c_{\downarrow\downarrow(3,S)}
\end{array}\right) = \left(\begin{array}{cc}
    t_{C} & 0 \\ 
    0 & t_{C}
\end{array}\right) \left(\begin{array}{cc}
   c_{\downarrow\uparrow(2,S)}  & 0 \\ 
    0 & c_{\downarrow\downarrow(2,S)}
\end{array}\right) + \left(\begin{array}{cc}
    t_{C} & 0 \\ 
    0 & t_{C} 
\end{array}\right) \left(\begin{array}{cc}
   c_{\downarrow\uparrow(4,S)}  & 0 \\ 
    0 & c_{\downarrow\downarrow(4,S)}
\end{array}\right) \nonumber \\
 + \left(\begin{array}{cc}
   \eta & 0 \\ 
    0 & \eta
\end{array}\right) \left(\begin{array}{cc}
    \tau_{\downarrow\uparrow(S\rightarrow B2)}e^{ik_{(B2)}a} \\
    \tau_{\downarrow\downarrow(S\rightarrow B2)}e^{ik_{(B2)}a} 
    \end{array}\right)\nonumber \\
\left[\left(\begin{array}{cc}
    E & 0 \\ 
    0 & E 
\end{array}\right) - \left(\begin{array}{cc}
    \epsilon & 0 \\  
    0 & \epsilon 
\end{array}\right)\right] \left(\begin{array}{cc}
   c_{\downarrow\uparrow(4,S)}  & 0 \\ 
    0 & c_{\downarrow\downarrow(4,S)}
\end{array}\right)
 = \left(\begin{array}{cc}
    t_{C} & 0 \\ 
    0 & t_{C}
\end{array}\right) \left(\begin{array}{cc}
   c_{\downarrow\uparrow(3,S)}  & 0 \\ 
    0 & c_{\downarrow\downarrow(3,S)}
\end{array}\right) +  \left(\begin{array}{cc}
    t_{D} & 0 \\ 
    0 & t_{D} 
\end{array}\right) \left(\begin{array}{cc}
    \tau_{\downarrow\uparrow(S\rightarrow D)}e^{ik_{(D)}a} \\
    \tau_{\downarrow\downarrow(S\rightarrow D)}e^{ik_{(D)}a} 
    \end{array}\right)\nonumber \\
\left[\left(\begin{array}{cc}
    E & 0 \\ 
    0 & E 
\end{array}\right) - \left(\begin{array}{cc}
    \epsilon_0 & 0 \\ 
    0 & \epsilon_0 
\end{array}\right)\right]
\left(\begin{array}{cc}
    \tau_{\downarrow\uparrow(S\rightarrow D)}e^{ik_{(D)}a} \\
    \tau_{\downarrow\downarrow(S\rightarrow D)}e^{ik_{(D)}a} 
    \end{array}\right) =  \left(\begin{array}{cc}
    t_{D} & 0 \\ 
    0 & t_{D} 
\end{array}\right) \left(\begin{array}{cc}
   c_{\downarrow\uparrow(4,S)}  & 0 \\ 
    0 & c_{\downarrow\downarrow(4,S)}
\end{array}\right) +
 \left(\begin{array}{cc}
    t_{0} & 0 \\
0 & t_{0}
\end{array}\right) \left(\begin{array}{cc}
    \tau_{\downarrow\uparrow(S\rightarrow D)}e^{2ik_{(D)}a} \\
    \tau_{\downarrow\downarrow(S\rightarrow D)}e^{2ik_{(D)}a}
    \end{array}\right)\nonumber \\
\left[\left(\begin{array}{cc}
    E & 0 \\
    0 & E
\end{array}\right) - \left(\begin{array}{cc}
    \epsilon_0 + \frac{2}{3}V_b & 0 \\
    0 & \epsilon_0 + \frac{2}{3}V_b
\end{array}\right)\right] \left(\begin{array}{cc}
    \tau_{\downarrow\uparrow(S\rightarrow B1)}e^{ik_{(B1)}a} \\
    \tau_{\downarrow\downarrow(S\rightarrow B1)}e^{ik_{(B1)}a}
    \end{array}\right)
 = \left(\begin{array}{cc}
   \eta & 0 \\
    0 & \eta
\end{array}\right) \left(\begin{array}{cc}
   c_{\downarrow\uparrow(2,S)}  & 0 \\
    0 & c_{\downarrow\downarrow(2,S)}
\end{array}\right)+ \left(\begin{array}{cc}
    t_0 & 0 \\
    0 & t_0
\end{array}\right) \left(\begin{array}{cc}
    \tau_{\downarrow\uparrow(S\rightarrow B1)}e^{2ik_{(B1)}a} \\
    \tau_{\downarrow\downarrow(S\rightarrow B1)}e^{2ik_{(B1)}a}
    \end{array}\right)\nonumber \\
\left[\left(\begin{array}{cc}
    E & 0 \\
0 & E
\end{array}\right) - \left(\begin{array}{cc}
    \epsilon_0 + \frac{1}{3}V_b & 0 \\
    0 & \epsilon_0 + \frac{1}{3}V_b
\end{array}\right)\right] \left(\begin{array}{cc}
    \tau_{\downarrow\uparrow(S\rightarrow B2)}e^{ik_{(B2)}a} \\
    \tau_{\downarrow\downarrow(S\rightarrow B2)}e^{ik_{(B2)}a}
    \end{array}\right)
 = \left(\begin{array}{cc}
   \eta & 0 \\
    0 & \eta
\end{array}\right) \left(\begin{array}{cc}
   c_{\downarrow\uparrow(3,S)}  & 0 \\
    0 & c_{\downarrow\downarrow(3,S)}
\end{array}\right) + \left(\begin{array}{cc}
    t_0 & 0 \\
    0 & t_0
\end{array}\right) \left(\begin{array}{cc}
    \tau_{\downarrow\uparrow(S\rightarrow B2)}e^{2ik_{(B2)}a} \\
    \tau_{\downarrow\downarrow(S\rightarrow B2)}e^{2ik_{(B2)}a}
    \end{array}\right). \nonumber \\
\label{eq14}
\end{eqnarray}}
\end{widetext}
The factors $\tau_{\sigma\sigma'}$ and $\rho_{\sigma\sigma'}$ used in 
these expressions correspond to the transmission and reflections amplitudes,
respectively, for an electron with spin $\sigma$ which is transmitted and/or
reflected as spin $\sigma'$.

\end{document}